\documentclass[twoside]{article}
\usepackage{graphicx}
\usepackage{fontenc}
\usepackage[cp1251]{inputenc}
\usepackage{fancyhdr}
\usepackage{cmpj2e}

\title[Exact solution of the mixed spin-$1/2$ and spin-$S$ Ising-Heisenberg diamond chain]
{Exact solution of the mixed spin-$1/2$ and spin-$S$ Ising-Heisenberg diamond chain}

\author[L. \v{C}anov\'a \textit{et al}.]{L. \v{C}anov\'a\refaddr{tuke}, J. Stre\v{c}ka\refaddr{upjs},
        and T. Lu\v{c}ivjansk\'y\refaddr{upjs}}
\addresses{
\addr{tuke} Department of Applied Mathematics, Faculty of Mechanical Engineering, \\
Technical University, Letn\'a 9, 042 00 Ko\v{s}ice, Slovak Republic \\
\addr{upjs} Department of Theoretical Physics and Astrophysics, Faculty of Science, \\
P. J. \v{S}af\'arik University, Park Angelinum 9, 040 01 Ko\v{s}ice, Slovak Republic}

\begin{document}

\maketitle

\begin{abstract}
The geometric frustration in a class of the mixed spin-$1/2$ and spin-$S$ Ising-Heisenberg 
diamond chains is investigated by combining three exact analytical techniques: Kambe projection 
method, decoration-iteration transformation and transfer-matrix method. The ground state, 
the magnetization process and the specific heat as a function of the external magnetic field 
are particularly examined for different strengths of the geometric frustration. It is shown that 
the increase of the Heisenberg spin value $S$ raises the number of intermediate magnetization 
plateaux, which emerge in magnetization curves provided that the ground state is highly degenerate 
on behalf of a sufficiently strong geometric frustration. On the other hand, all intermediate magnetization plateaux merge into a linear magnetization versus magnetic field dependence in the limit 
of classical Heisenberg spin $S \to \infty$. The enhanced magnetocaloric effect with cooling 
rate exceeding the one of paramagnetic salts is also detected when the disordered frustrated 
phase constitutes the ground state and the external magnetic field is small enough.

\keywords Ising-Heisenberg model, diamond chain, geometric frustration, exact results
\pacs 05.30.-d, 05.50.+q, 75.10.Hk, 75.10.Jm, 75.10.Pq, 75.40.Cx
\end{abstract}

\section{Introduction}

The quantum Heisenberg model with diamond chain topology has enjoyed a great scientific interest 
since two unusual tetramer-dimer and dimer-monomer phases were theoretically predicted by Takano \textit{et al}.~\cite{Tak96} in the zero-field ground-state phase diagram of the spin-$1/2$ 
Heisenberg diamond chain as a result of the mutual interplay between quantum fluctuations
and geometric frustration. Motivated by this discovery, several other one-dimensional (1D) 
quantum spin models consisting of diamond-shaped units have been suggested and solved with 
the aim to bring insight into a frustrated magnetism of diamond chain systems. It is now widely accepted that the zero-field ground-state phase diagram of the spin-$1/2$ Heisenberg model 
with the distorted diamond chain topology is rather complex and consists of the usual 
ferrimagnetic phase as well as several quantum dimerized and plaquette states \cite{Oka99,Wan02,Li08,Ton00,Gu07}. Further theoretical studies devoted to this 1D quantum spin model 
have provided accurate results for the ground-state phase diagram in a presence of the external magnetic field~\cite{Ton00,Gu07,Ton01,Oka03}, the spin gap~\cite{San00}, the magnetization and
susceptibility~\cite{Hon01}, as well as, the inversion phenomenon, which can be induced 
through the exchange anisotropy~\cite{Oka02,Oka05}. It is noteworthy that the ground state 
and thermodynamics of the mixed-spin diamond chains constituted by higher spins have been 
particularly examined as well~\cite{Nig97,Nig98,Pat03,Vit02a,Vit02b}.

It should be also mentioned, however, that the immense theoretical interest focused on 
the diamond chain structures is not purposeless. An important stimulus for a theoretical 
treatment of diamond chain models comes from rather recent findings that several insulating 
magnetic materials such as azurite Cu$_3$(CO$_3$)$_2$(OH)$_2$~\cite{Kik03,Oht03,Kik04,Oht04,Kik05a,Kik05b,Rul08}, Bi$_4$Cu$_3$V$_2$O$_{14}$~\cite{Sak02}, and Cu$_3$(TeO$_3$)$_2$Br$_2$~\cite{Uem07}, represent 
possible experimental realizations of the spin-$1/2$ Heisenberg diamond chain. It is convenient to remark that the natural mineral azurite was for a long time regarded as the best known candidate 
of the diamond chain compound even though recent ab-initio calculations indicate possibly non-negligible inter-chain interactions in this compound \cite{HFM08}. Notwithstanding this fact, 
experimental data measured for the azurite are in a good qualitative accordance with the relevant theoretical predictions for the highly frustrated spin-$1/2$ Heisenberg diamond chain. 
As a matter of fact, the magnetization plateau at one-third of the saturation magnetization~\cite{Oht03,Kik04,Oht04,Kik05a,Kik05b}, the double-peak structure in temperature 
dependences of the specific heat~\cite{Kik04,Kik05a,Kik05b,Rul08} and the zero-field susceptibility~\cite{Kik03,Oht03,Kik04,Kik05a,Kik05b} have been found both theoretically 
as well as experimentally. In addition, the spin-$1$, spin-$3/2$, and spin-$5/2$ Heisenberg model 
with the diamond chain topology might prove its usefulness in elucidating magnetic properties 
of polymeric coordination compounds M$_3$(OH)$_2$ (M = Ni, Co, Mn)~\cite{Gui02,Ton05}, [Ni$_3$(fum)$_2$($\mu_3$-OH)$_2$(H$_2$O)$_4$]$\cdot$(2H$_2$O)~\cite{Kon02}
and [Co$_3$(NC$_5$H$_3$(COO)$_2$)$_2$($\mu_3$-H$_2$O)$_2$(H$_2$O)$_2$]~\cite{Hum04}.

Unfortunately, the rigorous theoretical treatment of geometrically frustrated quantum Heisenberg 
models is very difficult to deal with due to a non-commutability of spin operators involved in 
the Heisenberg Hamiltonian, which is also a primary cause of a presence of quantum fluctuations. 
Owing to this fact, we have recently proposed a novel class of the geometrically 
frustrated Ising-Heisenberg diamond chain models \cite{Jas04,Can04,Can06,Str08}, which overcome 
this mathematical difficulty by introducing the Ising spins at the nodal sites and the Heisenberg dimers on the interstitial decorating sites of the diamond chain. This simplified quantum 
model can be examined within the framework of exact analytical approach based on the generalized decoration-iteration transformation~\cite{Fis59,Syo72,Roj09}, because the nodal Ising spins represent 
a barrier for quantum fluctuations that are consequently restricted to elementary diamond-shaped units. It is worth mentioning that the relatively simple analytical technique based on the generalized decoration-iteration transformation has been recently adapted to explore an effect of the geometric frustration also in the asymmetric Ising-Heisenberg tetrahedral chain~\cite{Val08}, the Ising-Heisenberg chain consisting of triangular-shaped Heisenberg trimers alternating with 
the nodal Ising spins~\cite{Ant09,Oha09}, as well as, the kinetically frustrated diamond chain 
models constituted by the nodal Ising spins and mobile electrons delocalized over the interstitial 
decorating sites \cite{Per08,Per09,Lis09}.

The main purpose of this work is to provide the exact solution for the generalized version of the 
mixed spin-$1/2$ and spin-$S$ Ising-Heisenberg diamond chain \cite{Jas04,Can04,Can06,Str08}, which should bring a deeper insight into how the magnetic properties depend on the quantum spin number $S$ of the Heisenberg spins. The exact analytical solution for this extended version of the Ising-Heisenberg diamond chain will be attained by combining the Kambe projection method~\cite{Kam50} with the generalized decoration-iteration mapping transformation \cite{Fis59,Syo72,Roj09} and the transfer-matrix technique \cite{Kra44,Bax82}.

The outline of this paper is as follows. In section~\ref{sec:Model}, we present at first a detailed formulation of the Ising-Heisenberg model, which is subsequently followed by a brief description of basic steps of exact analytical treatment. Section~\ref{sec:Results} deals with the interpretation of the most interesting results for the ground-state phase diagrams, the magnetization process, the specific heat, and the magnetocaloric effect. Finally, some concluding remarks are drawn in
section~\ref{sec:Conclusions}.

\section{Model and its exact solution}
\label{sec:Model}

\begin{figure}[h]
\vspace{-1.0cm}
\centerline{\includegraphics[width=0.45\textwidth]{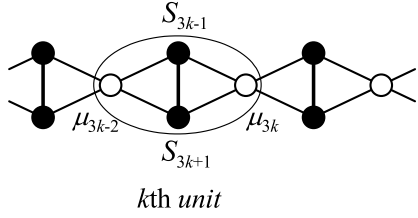}}
\vspace{-1.0cm}
\caption{A part of the mixed spin-1/2 and spin-$S$ Ising-Heisenberg diamond chain. The empty circles denote lattice positions of the Ising spins $\mu = 1/2$, while the filled circles label lattice positions of the Heisenberg spins of an arbitrary magnitude $S$. The ellipse demarcates spins 
belonging to the $k$th diamond unit.}
\label{fig1}
\end{figure}

Let us consider an one-dimensional lattice of inter-connected diamonds as schematically illustrated 
in figure~\ref{fig1}. In this figure, empty circles denote nodal lattice sites occupied by the 
\textit{Ising spins} $\mu = 1/2$, while the filled ones label interstitial (decorating) lattice sites occupied by the \textit{Heisenberg spins} with an arbitrary quantum spin number $S$. The total Hamiltonian of this mixed-spin Ising-Heisenberg diamond chain can be for further convenience written 
as a sum over cluster Hamiltonians $\hat{{\cal H}} = \sum_{k = 1}^{N} \hat{{\cal{H}}}_k$, where each cluster Hamiltonian $\hat{{\cal H}}_k$ involves all the interaction terms belonging to the $k$th 
diamond-shaped unit (see figure~\ref{fig1})
\begin{equation}
\label{eq:Hk}
\hat{\mathcal H}_k = J_{H} \vec{\bf S}_{3k-1} \cdot \vec{\bf S}_{3k}
                  + J_{I} ({\hat{S}^{z}}_{3k-1} + {\hat{S}^{z}}_{3k})
                              ({\hat{\mu}^{z}}_{3k-2} + {\hat{\mu}^{z}}_{3k+1})
                  - H_{H} ({\hat{S}^{z}}_{3k-1} + {\hat{S}^{z}}_{3k})
                  - H_{I} ({\hat{\mu}^{z}}_{3k-2} + {\hat{\mu}^{z}}_{3k+1})/2.
\end{equation}
Here, ${\hat{\mu}^{z}}_{k}$ and $\vec{\bf S}_{k} = ({\hat{S}^{x}}_{k}, {\hat{S}^{y}}_{k}, {\hat{S}^{z}}_{k})$ denote spatial components of the spin-$1/2$ and spin-$S$ operators,
the parameter $J_{H}$ labels the isotropic exchange interaction between the nearest-neighbouring Heisenberg spins and the parameter $J_{I}$ denotes the Ising interaction between the Heisenberg 
spins and their nearest Ising neighbours. Finally, the last two terms determine the magnetostatic Zeeman's energy of the Ising and Heisenberg spins placed in an external magnetic field $H_{I}$ 
and $H_{H}$ oriented along the $z$-axis, respectively.

The most important point of our calculation represents an evaluation of the partition function. 
Taking into account a validity of the commutation relation $[\hat{\mathcal{H}}_k, \hat{\mathcal{H}}_l] = 0$ between cluster Hamiltonians of two different diamond units, the partition function of the  Ising-Heisenberg model can be partially factorized into the product
\begin{equation}
\label{eq:Z}
{\cal{Z}} = \sum_{\{ \mu_k \}} {\rm Tr}_{\{ S_k \}} \exp(-\beta\hat{{\mathcal H}})
          = \sum_{\{ \mu_k \}} \prod_{k=1}^N {\rm Tr}_{k} \exp(-\beta\hat{{\mathcal H}_k}).
\end{equation}
Above, $\beta = 1/(k_{\rm B}T)$, $k_{\rm B}$ is Boltzmann's constant, $T$ is the absolute temperature, the symbols $\sum_{\{ \mu_k \}}$ and ${\rm Tr}_{\{ S_k \}}$ denote summations over spin degrees 
of freedom of all Ising and Heisenberg spins, respectively, and the symbol ${\rm Tr}_{k}$ stands 
for a trace over spin degrees of freedom of both Heisenberg spins from the $k$th diamond plaquette. 
It is quite obvious from the equation~(\ref{eq:Z}) that it is necessary to perform this latter
partial trace in order to proceed further with a calculation. The calculation of the partial trace 
can be easily accomplished with the help of Kambe projection method~\cite{Kam50}, since the cluster Hamiltonian $\hat{\mathcal H}_k$ can alternatively be viewed as the Hamiltonian of the spin-$S$ Heisenberg dimer placed in the effective field $H_{\it eff} = H_{H} - J_{I} (\mu^{z}_{3k-2} + \mu^{z}_{3k+1})$. Consequently, the complete set of eigenvalues $E_k$ corresponding to the cluster Hamiltonian (\ref{eq:Hk}) can be expressed solely in terms of two quantum spin numbers 
$S_{tot}$ and $S_{tot}^z$,
\begin{equation}
\label{eq:Ek}
 E_k (S_{tot}, S_{tot}^z)
= - J_{H} S(S + 1) + J_{H} S_{tot} (S_{tot} + 1)/2
  - H_{\it eff} S_{tot}^z  - H_{I} (\mu^{z}_{3k-2} + \mu^{z}_{3k+1})/2,
\end{equation}
which determine the total quantum spin number of the spin-$S$ Heisenberg dimer and its projection towards the $z$-axis, respectively. According to the basic laws of quantum mechanics, the total quantum spin number of the spin-$S$ dimer gains $2S+1$ different values $S_{tot} = 0, 1, \ldots, 2S$, while the quantum spin number $S_{tot}^z$ gains $2S_{tot}+1$ different values $S_{tot}^z = -S_{tot}, -S_{tot}+1, \ldots, S_{tot}$ for each allowed value of $S_{tot}$. Then, the energy eigenvalues (\ref{eq:Ek}) can 
be straightforwardly used to obtain the relevant trace emerging in the last expression on right-hand-side of the equation~(\ref{eq:Z}). In addition, the resulting expression immediately implies a possibility of performing the generalized decoration-iteration transformation~\cite{Fis59,Syo72,Roj09}
\begin{eqnarray}
\label{eq:DIT}
{\rm Tr}_{k} \exp(-\beta\hat{{\mathcal H}_k})
&=& \exp[\beta J_{H} S(S + 1) + \beta H_{I}(\mu^{z}_{3k-2} + \mu^{z}_{3k+1})/2]
\nonumber \\
& & \times \sum_{n=0}^{2S} \sum_{m=-n}^{n} \!\! \exp [-\beta J_{H}n(n + 1)/2]\cosh(\beta H_{\it eff} m)
\nonumber \\
&=& A \exp[\beta R \mu^{z}_{3k-2} \mu^{z}_{3k+1}
          + \beta H_{0} (\mu^{z}_{3k-2} + \mu^{z}_{3k+1})/2].
\end{eqnarray}
From the physical point of view, the mapping transformation (\ref{eq:DIT}) effectively removes 
all the interaction parameters associated with a couple of the Heisenberg spins from the $k$th 
diamond unit and replaces them by the effective interaction $R$ and the effective field $H_0$ 
acting on the remaining Ising spins $\mu_{3k-2}$ and $\mu_{3k+1}$ only. In this way, one establishes 
a simple connection between the exact solution of the mixed spin-$1/2$ and spin-$S$ Ising-Heisenberg diamond chain and the uniform spin-$1/2$ Ising linear chain with the effective nearest-neighbour interaction $R$ and the effective magnetic field $H_{0}$. Of course, the transformation relation (\ref{eq:DIT}) must hold for all possible spin combinations of the Ising spins $\mu_{3k-2}$ and
$\mu_{3k+1}$. This 'self-consistency' condition then unambiguously determines so far not specified mapping parameters $A$, $R$ and $H_{0}$,
\begin{equation}
\label{eq:AR}
A = \exp[\beta J_{H} S(S + 1)] (W_+W_-W^2)^{1/4}, \,\,\,
\beta R = \ln\!\left(\frac{W_{+}W_{-}}{W^2}\right)\!, \,\,\,
\beta H_{0} = \beta H_{I} - \ln\!\left(\frac{W_{+}}{W_{-}}\right)\!,
\end{equation}
which can be uniquely expressed through the functions $W_{\pm} = F(\pm 1)$ and $W = F(0)$ with \begin{equation}
\label{eq:F(x)}
F(x) = \sum_{n=0}^{2S} \sum_{m=-n}^{n} \!\!
       \exp [-\beta J_{H}n(n + 1)/2]\cosh[\beta m (J_{I} x + H_{H})].
\end{equation}
Now, a direct substitution of the transformation (\ref{eq:DIT}) into the expression (\ref{eq:Z}) 
yields the equality
\begin{equation}
\label{eq:ZihZ0}
{\cal Z} (\beta, J_{H}, J_{I}, H_{H}, H_{I}) = A^{N} {\cal Z}_{0}(\beta, R, H_{0}),
\end{equation}
which establishes an exact mapping relationship between the partition function $\cal{Z}$ of the mixed-spin Ising-Heisenberg diamond chain and the partition function ${\cal Z}_{0}$ of the uniform
spin-$1/2$ Ising linear chain with the nearest-neighbour coupling $R$ and the effective magnetic 
field $H_{0}$. It is valuable to remark that the mapping relation (\ref{eq:ZihZ0}) in fact 
completes our exact calculation of the partition function, since the partition function of 
the uniform spin-$1/2$ Ising chain can simply be calculated within the framework of the 
transfer-matrix method~\cite{Kra44,Bax82}
\begin{equation}
\label{eq:Z0}
{\cal Z}_{0}(\beta, R, H_{0}) = \exp(N \beta R/4) \left[ \cosh \left(\beta H_0/2 \right) 
+ \sqrt{\sinh^2 \left(\beta H_0/2 \right) + \exp(- \beta R)} \right]^N.
\end{equation}
At this stage, the mapping relationship (\ref{eq:ZihZ0}) between the partition functions can in turn  
be utilized also for a straightforward calculation of the Helmholtz free energy of the Ising-Heisenberg diamond chain
\begin{equation}
\label{eq:F}
{\cal F} = - k_{\rm B} T \ln {\cal Z} = {\cal{F}}_{0} - N k_{\rm B} T \ln A,
\end{equation}
where ${\cal F}_{0} = - k_{\rm B} T \ln {\cal Z}_0$ represents the Helmholtz free energy
of the corresponding spin-$1/2$ Ising chain. Subsequently, some other important physical 
quantities can readily be calculated by using the standard thermodynamic relations. Indeed, 
the sub-lattice magnetization $m_{I}$ and $m_{H}$ reduced per one Ising and Heisenberg spin, respectively, can easily be obtained by differentiating equation~(\ref{eq:F}) with respect 
to the particular magnetic field acting on Ising and Heisenberg spins
\begin{eqnarray}
m_{I} \!\!\!&=&\! - \frac{1}{N} \!\left( \frac{\partial
{\mathcal F}}{\partial H_{I}} \right)_{\!T} = - \frac{1}{N}
\!\left( \frac{\partial {\mathcal F}_{0}}{\partial
H_{0}}\right)_{\!T} = m_{0}\,, \label{eq:mI}
\\
m_{H} \!\!\!\!&=&\!\! - \frac{1}{2N} \!\left( \frac{\partial {\mathcal F}}{\partial H_{H}} \right)_{\!T} =
\frac{{\cal L}_{H_{H}}^{+}}{2}\!\left(\frac{1}{4} - m_0 + \varepsilon_0\right)\! +
\frac{{\cal L}_{H_{H}}^{-}}{2}\!\left(\frac{1}{4} + m_0 + \varepsilon_0\right)\! +
{\cal L}_{H_{H}}\!\left(\frac{1}{4} - \varepsilon_0\right)\!,
\label{eq:mH}
\end{eqnarray}
In above, the parameters $m_{0}$ and $\varepsilon_0$ denote the single-site magnetization and 
the correlation function between the nearest-neighbour spins of the uniform spin-$1/2$ Ising chain \cite{Kra44,Bax82}, which are unambiguously given by equations (\ref{eq:AR}) for the mapping parameters $R$ and $H_{0}$ through
\begin{eqnarray}
m_{0} \!\!\!&=& \! \!\! \frac{1}{2} \frac{\sinh \left( \beta H_0/2 \right)}
                       {\sqrt{\sinh^2 \left( \beta H_0/2 \right) + \exp \left(- \beta R \right)}}, \label{eq:m0} \\
\varepsilon_0 \!\!\!\!&=& \!\! \!\! m_0^2 + \frac{\cosh \left(\beta H_0/2 \right) - \sqrt{\sinh^2 \left( \beta H_0/2 \right) + \exp \left(- \beta R \right)}}{\cosh \left(\beta H_0/2 \right) + \sqrt{\sinh^2 \left( \beta H_0/2 \right) + \exp \left(- \beta R \right)}} \left(1 - m_0^2 \right).
\label{eq:e0}
\end{eqnarray}
Finally, the coefficients ${\cal L}_{x}^{\pm}$ and ${\cal L}_{x}$ emerging in equation~(\ref{eq:mH}) mark the expressions ${\cal L}_{x}^{\pm} = \frac{\partial}{\partial x} \ln W_{\pm}$ and ${\cal
L}_{x} = \frac{\partial}{\partial x} \ln W$, whose explicit forms are too cumbersome to write them 
down here explicitly (note that their explicit forms directly follow from the definitions given 
above after a straightforward but rather lengthy calculations). Now, the total magnetization 
normalized per one spin of the diamond chain can be calculated from the expression 
$m = (m_{I} + 2m_{H})/3$. Furthermore, the entropy and specific heat of the mixed-spin Ising-Heisenberg diamond chain can directly be calculated from the standard thermodynamical relations by computing the first and second temperature derivatives of the Helmholtz free energy (\ref{eq:F}). 
In this way, one attains the following rigorous results for the entropy and specific heat 
per one site of the investigated Ising-Heisenberg diamond chain
\begin{eqnarray}
\frac{{\cal S}}{3N} \!\!&=&\!\!
\frac{1}{3N}\ln {\cal Z}_{0} + \frac{1}{3}\ln A
- \frac{\beta}{3}[J_{H}S(S + 1) + m_0H_{I}]
\nonumber \\
&&{}
- \frac{\beta{\cal L}_{\beta}^{+}}{3} \!\left(\frac{1}{4} - m_0 + \varepsilon_0\right)
- \frac{\beta{\cal L}_{\beta}^{-}}{3}\!\left(\frac{1}{4} + m_0 + \varepsilon_0\right)
- \frac{\beta{\cal L}_{\beta}}{3}\!\left(\frac{1}{2} - 2\varepsilon_0\right)\!,
\label{eq:S}
\\
\frac{{\cal C}}{3Nk_{\rm B}} \!\!&=&\!\!
\frac{\beta^2{\cal L}_{\beta\beta}^{+}}{3} \!\left(\frac{1}{4} - m_0 + \varepsilon_0\right)
+ \frac{\beta^2{\cal L}_{\beta\beta}^{-}}{3}\!\left(\frac{1}{4} + m_0 + \varepsilon_0\right)
+ \frac{\beta^2{\cal L}_{\beta\beta}}{3}\!\left(\frac{1}{2} - 2\varepsilon_0\right)
\nonumber \\
&&{}
- \frac{\beta^2{\cal L}_{\beta}^{+}}{3} \!\left( \frac{\partial m_0}{\partial\beta} - \frac{\partial \varepsilon_0}{\partial\beta} \right)
+ \frac{\beta^2{\cal L}_{\beta}^{-}}{3} \!\left( \frac{\partial m_0}{\partial\beta} + \frac{\partial \varepsilon_0}{\partial\beta} \right)
- \frac{2\beta^2{\cal L}_{\beta}}{3} \frac{\partial \varepsilon_0}{\partial\beta}
+ \frac{\beta^2 H_{\rm I}}{3}\frac{\partial m_0}{\partial\beta},
\label{eq:C}
\end{eqnarray}
where ${\cal L}_{\beta\beta}^{\pm} = \frac{\partial^2}{\partial \beta^2} \ln W_{\pm}$ and ${\cal L}_{\beta\beta} = \frac{\partial^2}{\partial \beta^2} \ln W$. As one can see, both afore-listed quantities are expressed in terms of the well-known exact results for the partition function 
${\cal Z}_{0}$, single-site magnetization $m_0$ and nearest-neighbour correlation function $\varepsilon_0$ of the corresponding spin-$1/2$ Ising linear chain~\cite{Kra44,Bax82}. 
Our exact calculation of the entropy and specific heat are thus essentially completed by evaluating
the coefficients ${\cal L}_{\beta}^{\pm}$, ${\cal L}_{\beta}$, ${\cal L}_{\beta\beta}^{\pm}$, 
${\cal L}_{\beta\beta}$ and inverse temperature derivatives of $m_0$ and $\varepsilon_0$, 
which are not explicitly given here for brevity.

\section{Results and discussion}
\label{sec:Results}

In this part, let us proceed to a discussion of the most interesting results obtained for the mixed spin-$1/2$ and spin-$S$ Ising-Heisenberg diamond chain. Before doing so, however, let us make few remarks on a validity of analytical results presented in the preceding section. It should be at first pointed out that all obtained results are rather general as they hold for arbitrary quantum 
spin number $S$ of the Heisenberg spins and also independently of whether ferromagnetic or antiferromagnetic interactions $J_{H}$ and $J_{I}$ are considered. It is also noteworthy that some particular cases of the investigated model system have already been examined by the present authors 
in earlier papers \cite{Jas04,Can04,Can06,Str08}. More specifically, the present results reduce 
to those acquired for the Ising-Heisenberg diamond chains with two particular spin values $S = 1/2$ 
and $1$, which have been undertaken a rather detailed analysis in references \cite{Jas04,Can04,Can06}. Furthermore, the present model also contains the mixed spin-$1/2$ and spin-$3/2$ Ising-Heisenberg diamond chain as another particular limiting case, which has been a subject matter of our preliminary report \cite{Str08} revealing a series of intermediate plateaux in the magnetization process of this quantum spin chain. 

With this background, we will restrict ourselves in this paper to an analysis of another particular spin cases of the mixed-spin Ising-Heisenberg diamond chain with the aim to shed light on how the magnetic behaviour of the model under investigation depends on the quantum spin number $S$ of the Heisenberg spins. Namely, this systematic study should provide a deeper understanding of the role 
of quantum fluctuations in determining the overall magnetic behaviour, because the quantum effects should become less significant by increasing the quantum spin number $S$. As we are mainly interested in the examination of the geometric frustration, we will assume in what follows the antiferromagnetic character of both interaction parameters $J_{H}>0$ and $J_{I}>0$. Next, 
it is also convenient to rescale all interaction parameters with respect to the Ising-type 
interaction $J_{I}$, which will henceforth serve as the energy unit. In doing so, one effectively introduces the following set of dimensionless parameters: $\alpha = J_{H}/J_{I}$, 
$h = H_{I}/J_{I} = H_{H}/J_{I}$, and $t = k_{\rm B}T/J_{I}$, as describing a strength of 
the geometric frustration, the external magnetic field, and the temperature, respectively.

\subsection{Ground-state properties}
\label{subsec:Ground-state}

\begin{figure}[htb]
\vspace{-0.5cm}
\centerline{\includegraphics[width=0.9\textwidth]{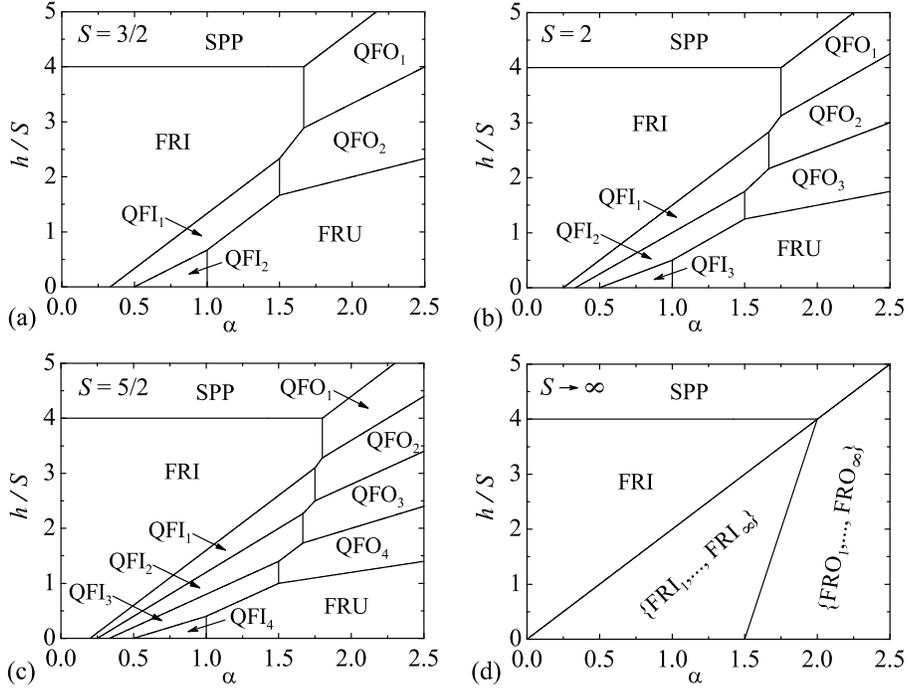}}
\vspace{-0.5cm}
\caption{Ground-state phase diagrams in the $\alpha-h$ plane
for the frustrated Ising-Heisenberg diamond chain with the
Heisenberg spin (a) $S = 3/2$, (b) $S = 2$, (c) $S = 5/2$
and (d) $S \to \infty$.}
\label{fig2}
\end{figure}

Let us begin our discussion by considering possible spin arrangements to emerge in the ground state 
of the mixed spin-$1/2$ and spin-$S$ Ising-Heisenberg diamond chain. For this purpose, the ground-state
phase diagrams in the $\alpha - h/S$ plane are displayed in figure~\ref{fig2} for four particular 
spin cases. As one can see from figures~\ref{fig2}(a)--(c), the mixed-spin diamond chain with 
finite quantum spin numbers $S = 3/2$, $2$, and $5/2$ have quite similar ground-state phase diagrams, which differ mainly in the total number of possible ground states. Apart from the semi-classically ordered ferrimagnetic phase (FRI), saturated paramagnetic phase (SPP) and frustrated phase (FRU), 
there also appear $2S-1$ quantum ferrimagnetic (to be denoted as QFI$_1$, QFI$_2$, \ldots, QFI$_{2S-1}$) and $2S-1$ quantum ferromagnetic (to be denoted as QFO$_1$, QFO$_2$, \ldots, QFO$_{2S-1}$) phases. Spin arrangements of the relevant phases can be unambiguously characterized 
by means of the following eigenfunctions and single-site magnetization
\begin{eqnarray}
|{\rm FRI} \rangle \!\!\!&=&\!\!\!  \prod_{k=1}^{N} | - \rangle_{3k-2}
\prod_{k=1}^{N} | S, S  \rangle_{3k-1, \, 3k}, \nonumber \\
&&m_{I} = -1/2,\, m_{H} = S,\,\, m/m_{sat}= 1 - 2/(4S+1);
\label{FRIa}
\\
|{\rm SPP} \rangle \!\!\!&=&\!\!\!  \prod_{k=1}^{N} | + \rangle_{3k-2}
\prod_{k=1}^{N} | S, S  \rangle_{3k-1, \, 3k}, \nonumber \\
&&m_{I} = 1/2,\, m_{H} = S,\,\, m/m_{sat}= 1;
\label{SPPa}
\\
|{\rm FRU} \rangle \!\!\!&=&\!\!\! \prod_{k=1}^{N} |\pm \rangle_{3k-2}
\prod_{k=1}^{N} \frac{1}{\sqrt{2S+1}} \sum_{l=-S}^{S}(-1)^{S+l}
|l, -l \rangle_{3k-1, \, 3k},
\nonumber \\
&&m_{I} = 0,\, m_{H} = 0,\,\, m/m_{sat}= 0;
\label{FRUa}
\\
|{\rm QFI}_{j} \rangle \!\!\!&=&\!\!\!  \prod_{k=1}^{N} | - \rangle_{3k-2}
\prod_{k=1}^{N} \sum_{l=S-j}^{S}(-1)^{S+l}A_{l, 2S-l-j}\,
|l, 2S-l-j \rangle_{3k-1, \, 3k}, \nonumber \\
&&m_{I} = -1/2,\, m_{H} = S-j/2,\,\, m/m_{sat}= 1 - 2(j+1)/(4S+1);
\label{QFIa}
\\
|{\rm QFO}_{\!j} \rangle \!\!\!&=&\!\!\!  \prod_{k=1}^{N} | + \rangle_{3k-2}
\prod_{k=1}^{N} \sum_{l=S-j}^{S}(-1)^{S+l}A_{l, 2S-l-j}\,
|l, 2S-l-j \rangle_{3k-1, \, 3k}, \nonumber \\
&&m_{I} = 1/2,\, m_{H} = S-j/2,\,\, m/m_{sat}= 1 - 2j/(4S+1);
\label{QFOa}
\end{eqnarray}
where $j = 1, 2, \ldots, 2S-1$. Note that the first product in the afore-listed eigenfunctions 
is taken over all Ising spins ($| \pm \rangle$ denotes $\mu^{z}=\pm1/2$), while the second one runs
over all Heisenberg dimers. The coefficients $A_{l, 2S-l-j}$ emerging in the last two eigenfunctions (\ref{QFIa}) and (\ref{QFOa}) represent probability amplitudes for finding the Heisenberg spin pairs 
in the spin state $|l, 2S-l-j \rangle$ and these are listed for several particular spin cases in the Appendix together with the complete analytic form of the eigenfunctions QFI$_{j}$ and QFO$_{j}$. Finally, $m_{sat}$ labels the saturation magnetization normalized 
per one spin of the diamond chain. 

It is quite evident from the set of equations (\ref{FRIa})--(\ref{QFOa}) that first two phases FRI 
and SPP exhibit spin arrangements, which are commonly observed also in the pure Ising systems. 
By contrast, the quantum entanglement between spin states of the Heisenberg spin pairs represents an inherent feature of the phases FRU, QFI$_j$ and QFO$_j$, which cannot be observed in the pure 
Ising systems. According to their location in the ground-state phase diagrams, the phases with entangled spin states of the Heisenberg spins appear as a result of a competition between the interaction parameters $\alpha$ and $h$, i.e., owing to a mutual interplay between the geometric frustration generated by the competition between the Heisenberg- and Ising-type interactions $J_{H}$ 
and $J_{I}$, respectively, the quantum fluctuations arising from the Heisenberg interaction $J_{H}$, 
and the effect of applied magnetic field $H$. It is also noteworthy that the phases QFI$_j$ and 
QFO$_j$ basically differ from each other just in a relevant spin alignment of the Ising spins, 
which are oriented antiparallel (parallel) with respect to the total spin of the Heisenberg dimers 
in QFI$_j$ (QFO$_j$). Hence, the overall spin arrangement of the phases QFI$_j$ (QFO$_j$) has 
typical features of the quantum ferrimagnetic (ferromagnetic) phase with a significant quantum reduction of the sublattice magnetization $m_{H}$ closely related to a quantum entanglement 
of the Heisenberg spins.

A profoundly different situation emerges when one is assuming the limit of classical Heisenberg spin 
$S \to \infty$, whose ground-state phase diagram is shown in figure~\ref{fig2}(d). As one can see, solid lines separate just four different phases, namely, the FRI phase, the SPP phase and two phases 
denoted as $\{$FRI$_1$, FRI$_2$, \ldots, FRI$_\infty \}$ and $\{$FRO$_1$, FRO$_2$, \ldots, 
FRO$_\infty \}$. As could be expected, the spin arrangements emerging in the phases FRI and SPP 
are very analogous to the ones of the diamond chains with finite values of the Heisenberg spins.
In the FRI phase, one actually finds $m_{I} = -1/2$, $m_{H} \to \infty$, $m/m_{sat}= 1$, which 
indicate the semi-classical ferrimagnetic character of this phase. In the SPP phase being stable at 
sufficiently strong fields, one observes $m_{I} = 1/2$, $m_{H} \to \infty$, $m/m_{sat}= 1$ 
implying that all Ising as well as Heisenberg spins are oriented in a direction of the applied 
external magnetic field. A detailed examination of another two possible phases $\{$FRI$_1$, FRI$_2$, \ldots, FRI$_\infty \}$ and $\{$FRO$_1$, FRO$_2$, \ldots, FRO$_\infty \}$ reveals that the former (latter) phase has ferrimagnetic (ferromagnetic) character, since the sublattice magnetization 
$m_{I} = -1/2$ ($1/2$) is oriented in the opposite (same) direction as the sublattice magnetization $m_{H}$, which linearly increases with the applied magnetic field $h$ on account 
of a gradual rotation of the classical Heisenberg spins into the external-field direction. 
The linear increase of the sublattice magnetization $m_{H}$ can alternatively be viewed also 
as a smooth sequence of infinite number of phase transitions between the phases FRI$_1$, 
FRI$_2$, $\ldots$, FRI$_\infty$ (or FRO$_1$, FRO$_2$, $\ldots$, FRO$_\infty$) representing 
classical analogs of the quantum phases QFI$_1$, QFI$_2$, $\ldots$, QFI$_\infty$ (or QFO$_1$, 
QFO$_2$, $\ldots$, QFO$_\infty$). However, each phase transition between those phases is connected 
with just an infinitesimal increase of the sublattice magnetization $m_{H}$ if $S \to \infty$ and hence, these field-induced transitions cannot be regarded as true phase transitions.

\subsection{Magnetization process}
\label{subsec:Mag.process}

\begin{figure}[htb]
\vspace{-0.8cm}
\centerline{\includegraphics[width=1.0\textwidth]{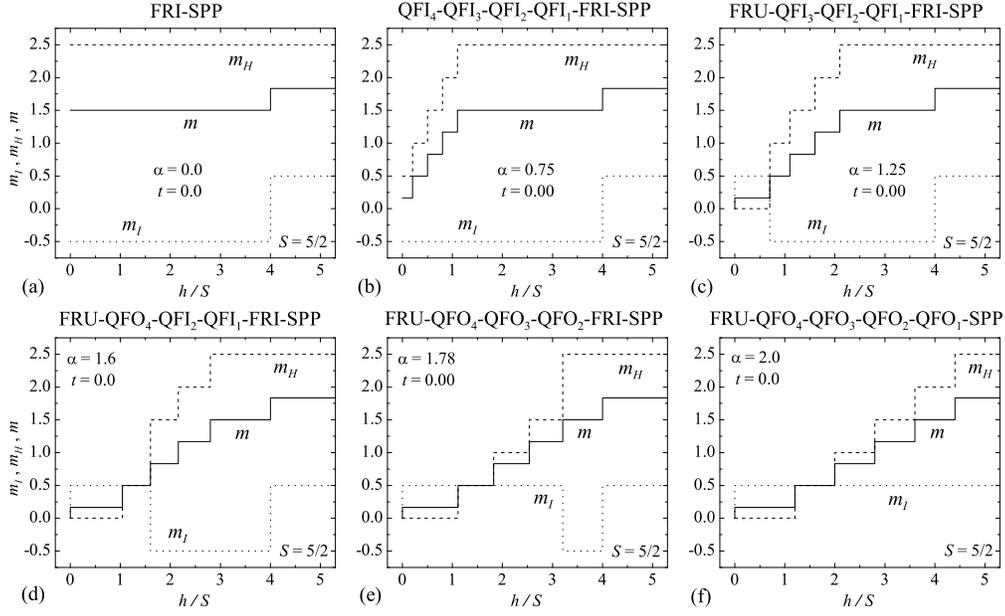}}
\vspace{-0.5cm}
\caption{The total and sublattice magnetization against the external magnetic field at zero
temperature $t = 0.0$ for the particular spin case $S = 5/2$ and the frustration parameter: 
(a) $\alpha = 0.0$, (b) $\alpha = 0.75$, (c) $\alpha = 1.25$, (d) $\alpha = 1.6$, 
(e) $\alpha = 1.78$, and (f) $\alpha = 2.0$.}
\label{fig3}
\end{figure}

Now, let us turn our attention to the magnetization process of the investigated model system.
It is useful to mention, however, that possible magnetization scenarios for the Ising-Heisenberg diamond chain with three lowest possible spin values of the Heisenberg spins $S = 1/2$, $1$ and $3/2$ have already been particularly examined in our previous papers~\cite{Can06,Str08} to which the interested reader is referred to for more details. In this respect, we merely depict in Fig.~\ref{fig3}
the total and sublattice magnetization for another particular case $S=5/2$, which should bring insight into how possible magnetization scenarios evolve by increasing significantly a magnitude of the Heisenberg spins. One may immediately come to the following general conclusions valid for the Ising-Heisenberg diamond chains with the arbitrary finite value $S$ of the Heisenberg spins by combining the formerly published results [see figures~2(a), 3, 7(a), 8 in reference~\cite{Can06} and figure~1 in reference~\cite{Str08}] with the results shown in figures~\ref{fig2}(a)--(c) and \ref{fig3}: there always exists the critical value of the frustration parameter $\alpha_{c} 
= (2S)^{-1}$ below which just a single intermediate magnetization plateau appears in the magnetization curve and above which more diverse magnetization scenarios can be in principle observed with two, three,$\ldots$, $2S$ intermediate magnetization plateaux. The zero-temperature magnetization 
curves corresponding to the particular case with $\alpha < \alpha_{c}$ essentially reflect the field-induced phase transition FRI-SPP, while they might reflect another five different sequences 
of phase transitions if $\alpha > \alpha_{c}$, namely,
\begin{enumerate}
    \item[1.] QFI$_{j}$-QFI$_{j-1}$- $\cdots$ -QFI$_{1}$-FRI-SPP ($j$ = 1,2,\ldots,2$S$-1)
    \item[2.] FRU-QFI$_{2S-2}$-QFI$_{2S-3}$- $\cdots$ -QFI$_{1}$-FRI-SPP
    \item[3.] FRU-QFO$_{2S-1}$-QFO$_{2S-2}$- $\cdots$ -QFO$_{j}$-QFI$_{j-2}$-QFI$_{j-3}$- $\cdots$  
              -QFI$_{1}$-FRI-SPP ($j$ = 3,4,\ldots,2$S$-1)
    \item[4.] FRU-QFO$_{2S-1}$-QFO$_{2S-2}$- $\cdots$ -QFO$_{2}$-FRI-SPP
    \item[5.] FRU-QFO$_{2S-1}$-QFO$_{2S-2}$- $\cdots$ -QFO$_{1}$-SPP
\end{enumerate}
Moreover, it is also worthwhile to remark that all intermediate plateaux identified in the relevant field dependences of the total magnetization appear at $1/(4S+1), 3/(4S+1), \ldots, (4S-1)/(4S+1)$ of the saturation magnetization. The closer mathematical analysis reveals that all these fractional values satisfy the Oshikawa-Yamanaka-Affeck rule $p\,(S_{u} - m) \in N$~\cite{Osh97}, which has been proposed as a necessary condition for the formation of quantized plateaux ($p$ is a period of the ground state, $S_u$ and $m$ are the total spin and total magnetization of the elementary unit).

\begin{figure}[htb]
\vspace{-0.8cm}
\centerline{\includegraphics[width=0.9\textwidth]{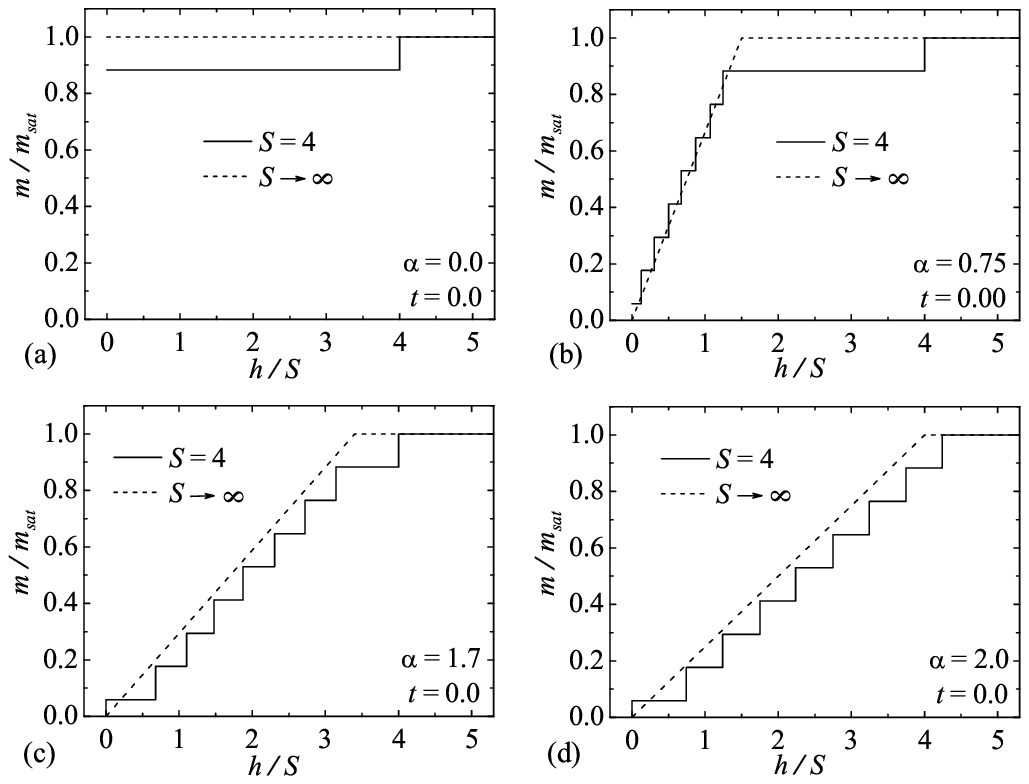}}
\vspace{-0.5cm}
\caption{The total magnetization reduced with respect to its saturation value versus the external field at zero temperature $t = 0.0$ for the spin cases $S = 4$ and $S \to \infty$ and the frustration parameter: (a) $\alpha = 0.05$, (b) $\alpha = 0.75$, (c) $\alpha = 1.7$, and (d) $\alpha = 2.0$.}
\label{fig4}
\end{figure}

To compare the magnetization scenario of the diamond chain with the finite and infinite value of the Heisenberg spin, typical magnetization versus magnetic field dependences are depicted in figure~\ref{fig4} for two particular spin cases $S = 4$ and $S \to \infty$. Note that the magnetization curves corresponding to the limit of classical Heisenberg spin $S \to \infty$ are for clarity shown as broken lines. It is easy to observe from this figure that the diamond chain with $S \to \infty$ has qualitatively different magnetization process in comparison with its quantum counterpart with the finite (albeit very high) value of the Heisenberg spins. Actually, the magnetization curves display a linear increase of the magnetization with increasing magnetic field until the magnetization reaches its saturation value if one considers the classical limit $S \to \infty$ instead of the sharp stepwise magnetization curves, which appear for any finite value of the Heisenberg spin. The observed linear increase of the magnetization in the phases $\{$FRI$_1$, FRI$_2$, \ldots,FRI$_\infty \}$ and 
$\{$FRO$_1$, FRO$_2$, \ldots, FRO$_\infty \}$
is consistent with the fact that the intermediate magnetization plateaux gradually shrink as the quantum spin number $S$ increases until they entirely merge together into the linear magnetization 
vs. external field dependence in the classical limit $S \to \infty$.

\subsection{Specific heat}
\label{subsec:Specific.heat}

\begin{figure}[htb]
\vspace{-0.7cm}
\centerline{\includegraphics[width=0.9\textwidth]{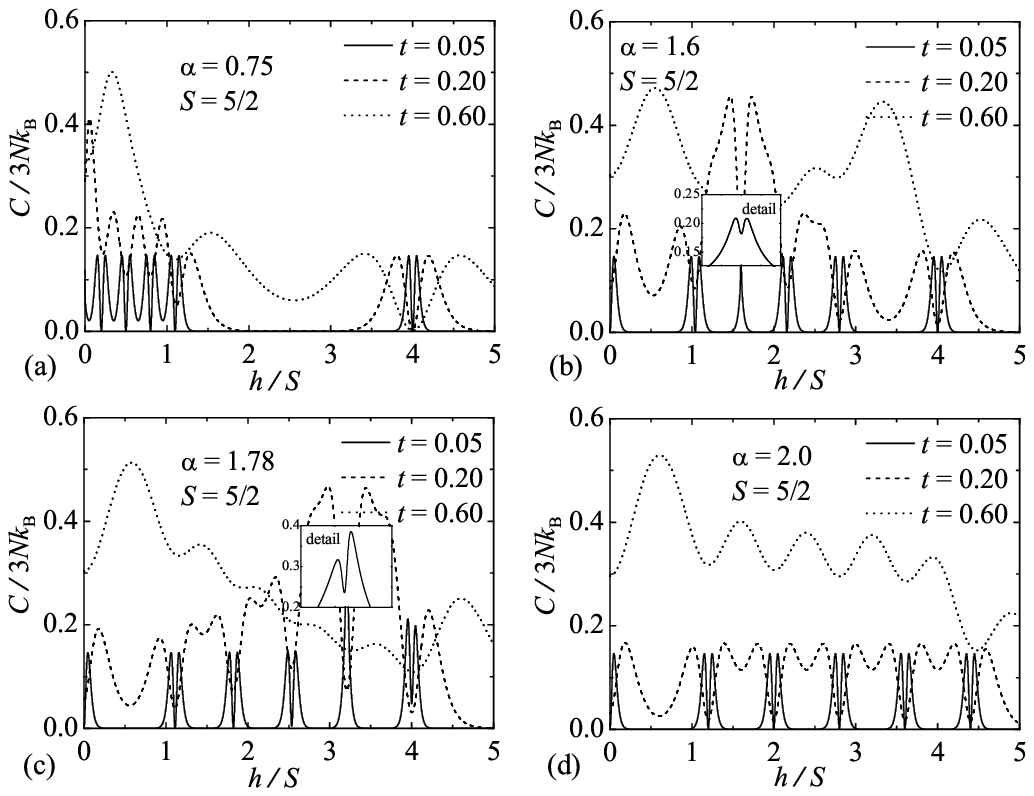}}
\vspace{-0.5cm}
\caption{The specific heat versus the external magnetic field at various temperatures $t = 0.05$, $0.2$, $0.6$ for the spin case $S = 5/2$ and the frustration parameter: (a) $\alpha = 0.75$, (b) $\alpha = 1.6$, (c) $\alpha = 1.78$, and (d) $\alpha = 2.0$. The details show the non-symmetric double peaks in the enlargened scale (for details see the text).}
\label{fig5}
\end{figure}

In this part, let us take a closer look at the specific heat versus external field dependences. 
Some typical variations of the specific heat as a function of the dimensionless 
magnetic field divided by the Heisenberg spin $S$ are depicted in figure~\ref{fig5} for the 
particular spin case $S = 5/2$, several values of the frustration parameter $\alpha$ and three different temperatures. To enable a direct comparison, the frustration parameter $\alpha$ 
is chosen so as to match four different magnetization scenarios plotted in figures~\ref{fig3}
(b), (d)-(f). The displayed sets of $C(h)$ curves thus basically reflect four magnetization processes, 
which are accompanied by the following sequences of the field-induced phase transitions: QFI$_{4}$-QFI$_{3}$-QFI$_{2}$-QFI$_{1}$-FRI-SPP [figure~\ref{fig5}(a)], FRU-QFO$_{4}$-QFI$_{2}$-QFI$_{1}$-FRI-SPP [figure~\ref{fig5}(b)], FRU-QFO$_{4}$-QFO$_{3}$-QFO$_{2}$-FRI-SPP [figure~\ref{fig5}(c)], and FRU-QFO$_{4}$-QFO$_{3}$-QFO$_{2}$-QFO$_{1}$-SPP [figure~\ref{fig5}(d)]. 
It can be clearly seen from these figures that the thermal trend of the specific heat
as a function of the magnetic field is quite similar for all investigated magnetization scenarios: $C(h)$ curves have rather irregular shapes with several broad maxima at higher temperatures (see the curves labeled as $t = 0.6$) that develop into more pronounced peaks gradually moving towards transition fields as the temperature is lowered (see for instance the curves $t = 0.2$). Naturally, 
the total number of peaks in the low-temperature $C(h)$ curves depends on the value of the frustration parameter $\alpha$. More specifically, the zero-temperature magnetization process formed 
by the sequence of field-induced transitions between the quantum ferrimagnetic phases [such as QFI$_{4}$-QFI$_{3}$-QFI$_{2}$-QFI$_{1}$-FRI-SPP shown in figure~\ref{fig5}(a)] characterize 
at sufficiently low temperatures identical double peaks in $C(h)$ curves being symmetrically 
centered around the respective transition fields.
On the other hand, one additional single peak is always detected in the low-temperature $C(h)$ 
curves in the limit of vanishing external field whenever the frustration parameter 
$\alpha$ drives the system into the disordered FRU ground state [see the case $t = 0.05$ in figures~\ref{fig5}(b)--(d)]. The origin of this single peak lies in thermal excitations of the frustrated Ising spins, which have tendency to align towards the external-field direction at 
low enough temperatures. Another interesting observation coming from figure \ref{fig5} is that 
almost all peaks are identical both in their height as well as width at low temperatures. 
The only exceptions constitute double peaks located around the critical fields associated 
with the phase transitions QFO$_4$-QFI$_{2}$ and QFO$_{2}$-FRI [see insets in figures~\ref{fig5}(b) and~(c)], whose origin significantly differs in some respects from the origin of others. 
As a matter of fact, both afore-mentioned double peaks result from intensive thermal excitations 
of Ising as well as Heisenberg spins, because both of them are changing their spin state at the transition fields between the phases QFO$_j$-QFI$_{j-2}$ ($j=1,2,\ldots,2S-1$) and QFO$_{2}$-FRI. Contrary to this, the other peaks reflect vigorous thermal excitations of only the Ising spins 
or only the Heisenberg spins as it directly follows from a detailed investigation 
of the low-temperature magnetization process shown in figure \ref{fig3}.

At this point, it is worthwhile to compare our results with those obtained by Efremov and Klemm 
for the antiferromagnetic spin-$S$ Heisenberg dimer \cite{Efr02,Efr06}. These authors have found 
that the low-temperature $C(h)$ curves of the antiferromagnetic spin-$S$ Heisenberg dimer exhibit 
a quite universal dependence with several marked double peaks symmetrically centered around level-crossing fields, which arise from thermal excitations between the ground state and first 
excited state. The relevant expression for the specific 
heat substantially simplifies in a close vicinity of the level-crossing fields at low temperatures, where the specific heat is well approximated by the equation (36) of the reference \cite{Efr06}, because the only significant contribution to the specific heat comes from thermal excitations between 
the ground state and the first excited state. In this respect, the specific heat goes exponentially 
to zero as one reaches the level-crossing field and there also appear two symmetrically centered 
peaks around the level-crossing field whose heights are given by the condition $C^{peak}/Nk_{\rm B} = (c/\!\cosh c)^2 \approx 0.439229$ ($c \approx 1.199679$ is the solution of the transcendent equation $\tanh c = 1/c$). It is noteworthy that the specific heat of the mixed-spin Ising-Heisenberg diamond chain is obviously governed by the same asymptotic expression near the transition fields as evidenced by the height and position of the relevant peaks in the field-dependence of the specific heat. 
In addition, it is also quite interesting to mention that the single peak, which is detected 
in the low-field tail ($h \to 0$) of the specific heat on assumption that the Ising–Heisenberg 
diamond chain is driven by the sufficiently strong frustration $\alpha >1$ towards the disordered FRU phase [see figures~\ref{fig5}(b)--(d)], is of the same height as the symmetrically centered double peaks. It is worthwhile to recall that this single peak originates from the field-induced splitting 
of the highly-degenerate lowest energy level and the same height of this peak indicates considerable contribution of the single spin-flip excitations of the Ising spins to the specific heat. 
On the other hand, the nonuniform double peaks observed in low-temperature $C(h)$ curves of the mixed-spin Ising-Heisenberg diamond chains cannot be described by the simple formula proposed 
by the analysis of the antiferromagnetic spin-$S$ Heisenberg dimer \cite{Efr06}. This implies 
that thermal excitations between more than two energy levels might possibly come into play 
at these particular transition fields, where both the Ising spins as well as the Heisenberg spins 
change their spin state.  

\subsection{Enhanced magnetocaloric effect}
\label{subsec:Magnetocaloric.effect}

Finally, we will turn our attention to an investigation of the adiabatic demagnetization studied 
in connection with the enhanced magnetocaloric effect.
\begin{figure}[htb]
\vspace{-0.7cm}
\centerline{\includegraphics[width=0.9\textwidth]{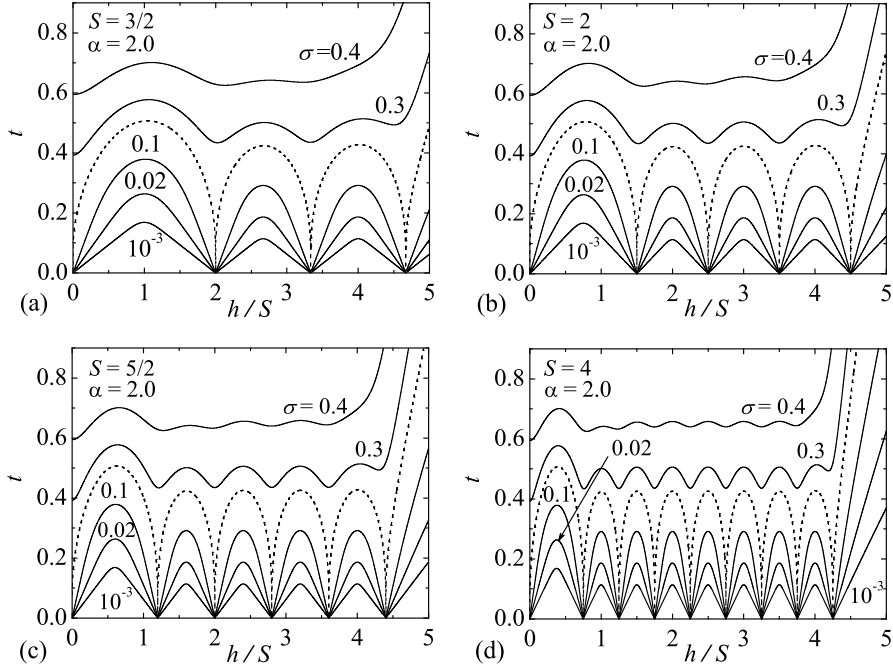}}
\vspace{-0.5cm}
\caption{Adiabatic demagnetization in the form of temperature versus external magnetic field 
dependence for several values of the entropy per one spin $\sigma = {\cal S}/3N$, the frustration 
ratio $\alpha= 2.0$ and the Heisenberg spins: (a) $S = 3/2$, (b) $S = 2$, (c) $S = 5/2$, 
and (d) $S = 4$. For better orientation, broken curves depict the relevant dependences when 
the entropy is equal to the residual value $\sigma_{res} = \frac13 \ln 2 \doteq 0.231049$.}
\label{fig6}
\end{figure}
For this purpose, the most interesting results for the field variations of the temperature are plotted 
in figure~\ref{fig6} for four different values of the Heisenberg spins $S$ by keeping the entropy constant at the particular value of the frustration parameter $\alpha = 2.0$. Note that this particular value of the frustration parameter was chosen so that the highly-degenerate FRU phase will constitute the ground state in the limit of vanishing external magnetic field. Namely, it has been 
recently proved that one achieves a substantial enhancement of the cooling rate during the adiabatic demagnetization of highly frustrated spin systems in comparison with the adiabatic demagnetization 
of unfrustrated spin systems \cite{Zhi03,Zhi04,Hon05,Der06,Hon06,Sch07}. As one can readily see 
from figures~\ref{fig5}(a)--(d), the mixed-spin Ising-Heisenberg diamond chains exhibit 
a pronounced valley-peak structure in the field dependence of the temperature at a fixed value 
of the entropy, which differ mainly in the total number of peaks (valleys) that gradually 
increases with the Heisenberg spin $S$. The most obvious drop (grow) of the temperature can always 
be found in the vicinity of zero field and transition fields at which the system undergoes zero-temperature phase transitions (see figure \ref{fig2} for the ground-state phase diagram). 
It can be also observed from figure \ref{fig6} that a cooling down to the lowest temperatures 
is achieved only if the entropy is less than or equal to the residual entropy 
$\sigma_{res} = {\cal S}_{res}/3N = \frac13 \ln 2$ of the disordered FRU phase, otherwise, 
the nonzero temperatures are finally reached as the external magnetic field vanishes. 

\begin{figure}[htb]
\vspace{-0.7cm}
\centerline{\includegraphics[width=0.9\textwidth]{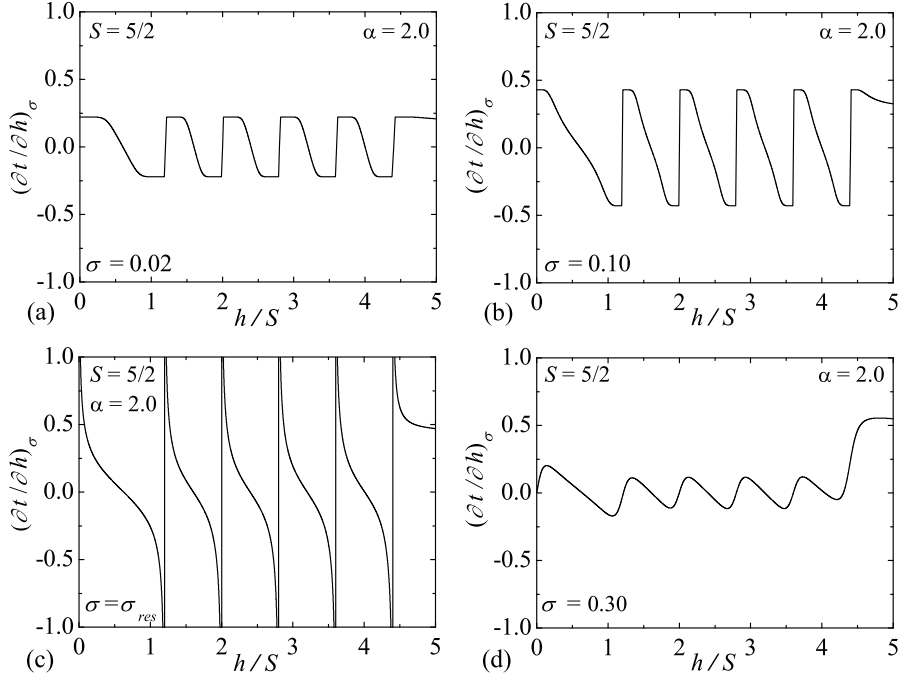}}
\vspace{-0.5cm}
\caption{Adiabatic magnetocaloric rate $(\partial t/\partial h)_{\sigma}$ as a function of the 
external magnetic field for the particular case with the Heisenberg spins $S=5/2$, the frustration ratio $\alpha= 2.0$, and four different values of the entropy per one spin ($\sigma = {\cal S}/3N$): (a) $\sigma = 0.02$, (b) $\sigma = 0.10$, (c) $\sigma_{res} = \frac13 \ln 2 \doteq 0.231049$, 
and (d) $\sigma = 0.30$.}
\label{fig7}
\end{figure}

To analyze the cooling rate of the adiabatic demagnetization in a more detail, the adiabatic magnetocaloric rate $(\partial t/\partial h)_{{\sigma}}$ is depicted in figure \ref{fig7} 
against the external magnetic field for the particular spin case $S=5/2$ and the frustration ratio $\alpha = 2.0$ by keeping the entropy per one spin $\sigma = {\cal S}/3N$ constant at 
four different values. As one can see, the most rapid cooling and heating rate can be observed 
during the adiabatic demagnetization when the entropy is close enough to the residual entropy 
$\sigma_{res} = \frac13 \ln 2$ of the disordered FRU phase. 
In the particular case $\sigma = \sigma_{res}$, one even observes asymptotically infinitely 
fast cooling (heating) as some particular transition field is approached from above (below). 
It is worthy of notice here that a relatively fast cooling of the frustrated diamond chain, 
which is observable in a certain range of fields and temperatures, might be of practical importance 
for a cooling purpose. The cooling effect is of technological relevance only if the relevant 
cooling rate exceeds the one of paramagnetic salts $(\partial t/\partial h)_{\sigma}^{\rm para} = t/h$ 
and the applied external magnetic fields are within the experimentally accessible range. 
From this point of view, the adiabatic demagnetization of the frustrated diamond chains becomes 
fairly efficient in producing the enhanced magnetocaloric effect just if the entropy is chosen 
close enough to its residual value $\sigma_{res}$ and the applied magnetic field is sufficiently 
small. This makes from the frustrated diamond chain systems promising candidates for being 
efficient refrigerant materials, which would allow reaching ultra-low temperatures unlike the paramagnetic salts that usually exhibit a spin-glass transition. However, the cooling rate 
basically diminishes when the entropy is selected far from the residual value $\sigma_{res}$ 
and this may represent rather inconvenient property hindering from real-world applications 
by reaching the ultra-low temperatures with the help of the adiabatic demagnetization.

\section{Conclusions}
\label{sec:Conclusions}

In the present article, the mixed spin-$1/2$ and spin-$S$ Ising-Heisenberg diamond chain has exactly been solved by combining three precise analytical techniques: the Kambe projection method, the generalized decoration-iteration transformation, and the transfer-matrix method. Within the framework of this exact analytical approach, the ground-state phase diagrams and the magnetization process 
were particularly examined in dependence on a strength of the geometric frustration, as well as, 
the quantum spin number $S$ of the Heisenberg spins. In addition, the specific heat as a function 
of external magnetic field has been explored in detail along with the adiabatic demagnetization process, which was examined in connection with the enhanced magnetocaloric effect.

The main goal of the present work was to shed light on how the magnetic behaviour of the mixed-spin diamond chains depends on a magnitude of the Heisenberg spins. The rather systematic study of the Ising-Heisenberg diamond chain models with different values of the Heisenberg spins indeed brought 
a deeper insight into how a diversity of the magnetization scenarios is related to the magnitude 
of the Heisenberg spins. It has been shown that the mixed spin-1/2 and spin-$S$ Ising-Heisenberg diamond chains generally exhibit multistep magnetization curves with up to $2S$ intermediate magnetization plateaux provided that there is a sufficiently strong geometric frustration. 
Beside this, we have also performed the comparative study of the classical limit 
of the Ising-Heisenberg diamond chain with the classical vector spins $S \to \infty$, which enabled 
us to discern the typical quantum features (e.g. multiply steps in the magnetization curves) 
from the typical classical ones (e.g. the linear magnetization vs. external field dependence).  

\section*{Acknowledgements}
\hfill\strut%
\\
This work was financially supported by the Slovak Research and Development Agency under the contract LPP-0107-06 and by Ministry of Education of SR under the grant No. VEGA 1/0128/08.

\section*{Appendix}
This appendix explicitly enumerates the coefficients $A_{l,2S-l-j}$, which emerge in the eigenfunctions (\ref{QFIa}) and (\ref{QFOa}) characterizing the quantum phases QFI$_{j}$ and QFO$_{j}$ ($j=1,2, \ldots, 2S-1$), that determine probability amplitudes for finding the Heisenberg spin pairs in the spin state $|l, 2S-l-j \rangle$. In what follows, the relevant probability amplitudes are listed for 
three particular spin values $S=1$, $3/2$ and $2$ of the Heisenberg spins together with the explicit form of the phases QFI$_{j}$ and QFO$_{j}$.

\begin{enumerate}
\item[1.] Heisenberg spin $S=1$: \\ 
\begin{eqnarray}
A_{1,0} \!\!\!&=&\!\!\! A_{0,1} =  \frac{1}{\sqrt{2}}; \nonumber \\ 
|{\rm QFI}_{1} \rangle \!\!\!&=&\!\!\!  \prod_{k=1}^{N} | - \rangle_{3k-2}
\prod_{k=1}^{N} \frac{1}{\sqrt{2}} \left( |1, 0 \rangle - |0, 1 \rangle \right)_{3k-1, \, 3k}; \nonumber \\
|{\rm QFO}_{1} \rangle \!\!\!&=&\!\!\!  \prod_{k=1}^{N} | + \rangle_{3k-2}
\prod_{k=1}^{N} \frac{1}{\sqrt{2}} \left( |1, 0 \rangle - |0, 1 \rangle \right)_{3k-1, \, 3k}. \nonumber 
\end{eqnarray}
   
\item[2.] Heisenberg spin $S=3/2$: \\ 
\begin{eqnarray}
A_{\frac{3}{2},\frac{1}{2}} \!\!\!&=&\!\!\! A_{\frac{1}{2},\frac{3}{2}} =  \frac{1}{\sqrt{2}};
A_{\frac{3}{2},-\frac{1}{2}} = A_{-\frac{1}{2},\frac{3}{2}} =  \sqrt{\frac{3}{10}}; 
A_{\frac{1}{2},\frac{1}{2}} = \sqrt{\frac{2}{5}};
\nonumber \\ 
|{\rm QFI}_{1} \rangle \!\!\!&=&\!\!\!  \prod_{k=1}^{N} | - \rangle_{3k-2}
\prod_{k=1}^{N} \frac{1}{\sqrt{2}} \left( \biggl| \frac{3}{2},\frac{1}{2} \biggr \rangle 
- \biggl| \frac{1}{2},\frac{3}{2} \biggr \rangle \right)_{3k-1, \, 3k}; \nonumber \\
|{\rm QFO}_{1} \rangle \!\!\!&=&\!\!\!  \prod_{k=1}^{N} | + \rangle_{3k-2}
\prod_{k=1}^{N} \frac{1}{\sqrt{2}} \left( \biggl| \frac{3}{2},\frac{1}{2} \biggr \rangle 
- \biggl| \frac{1}{2},\frac{3}{2} \biggr \rangle \right)_{3k-1, \, 3k};
\nonumber \\ 
|{\rm QFI}_{2} \rangle \!\!\!&=&\!\!\!  \prod_{k=1}^{N} | - \rangle_{3k-2} \prod_{k=1}^{N} \left[\sqrt{\frac{2}{5}} \biggl| \frac{1}{2},\frac{1}{2} \biggr \rangle - \sqrt{\frac{3}{10}} \left( \biggl| \frac{3}{2},-\frac{1}{2} \biggr \rangle + \biggl| -\frac{1}{2},\frac{3}{2} \biggr \rangle 
\right) \right]_{3k-1, \, 3k}; 
\nonumber \\
|{\rm QFO}_{2} \rangle \!\!\!&=&\!\!\!  \prod_{k=1}^{N} | + \rangle_{3k-2} \prod_{k=1}^{N} \left[\sqrt{\frac{2}{5}} \biggl| \frac{1}{2},\frac{1}{2} \biggr \rangle - \sqrt{\frac{3}{10}} \left( \biggl| \frac{3}{2},-\frac{1}{2} \biggr \rangle + \biggl|-\frac{1}{2},\frac{3}{2} \biggr \rangle \right) \right]_{3k-1, \, 3k}.  \nonumber
\end{eqnarray}

\item[3.] Heisenberg spin $S=2$: \\ 
\begin{eqnarray}
A_{2,1} \!\!\!&=&\!\!\! A_{1,2} = \frac{1}{\sqrt{2}}; A_{2,0} = A_{0,2} = \sqrt{\frac{2}{7}}; 
A_{1,1} = \sqrt{\frac{3}{7}}; A_{2,-1} = A_{-1,2} = \frac{1}{\sqrt{5}}; 
A_{1,0} = A_{0,1} = \sqrt{\frac{3}{10}}; 
\nonumber \\ 
|{\rm QFI}_{1} \rangle \!\!\!&=&\!\!\!  \prod_{k=1}^{N} | - \rangle_{3k-2}
\prod_{k=1}^{N} \frac{1}{\sqrt{2}} \left( |2, 1 \rangle - |1, 2 \rangle \right)_{3k-1, \, 3k}; \nonumber \\
|{\rm QFO}_{1} \rangle \!\!\!&=&\!\!\!  \prod_{k=1}^{N} | + \rangle_{3k-2}
\prod_{k=1}^{N} \frac{1}{\sqrt{2}} \left( |2, 1 \rangle - |1, 2 \rangle \right)_{3k-1, \, 3k}; \nonumber \\
|{\rm QFI}_{2} \rangle \!\!\!&=&\!\!\!  \prod_{k=1}^{N} | - \rangle_{3k-2}
\prod_{k=1}^{N} \left[ \sqrt{\frac{2}{7}} \left( |2, 0 \rangle + |0, 2 \rangle \right) 
- \sqrt{\frac{3}{7}} |1, 1 \rangle \right]_{3k-1, \, 3k}; 
\nonumber \\
|{\rm QFO}_{2} \rangle \!\!\!&=&\!\!\!  \prod_{k=1}^{N} | + \rangle_{3k-2}
\prod_{k=1}^{N} \left[ \sqrt{\frac{2}{7}} \left( |2, 0 \rangle + |0, 2 \rangle \right) 
- \sqrt{\frac{3}{7}} |1, 1 \rangle \right]_{3k-1, \, 3k}; 
\nonumber \\
|{\rm QFI}_{3} \rangle \!\!\!&=&\!\!\!  \prod_{k=1}^{N} | - \rangle_{3k-2}
\prod_{k=1}^{N} \left[ \frac{1}{\sqrt{5}} \left( |2, -1 \rangle - |-1, 2 \rangle \right) 
- \sqrt{\frac{3}{10}} \left( |1, 0 \rangle - |0, 1 \rangle \right) \right]_{3k-1, \, 3k}; 
\nonumber \\
|{\rm QFO}_{3} \rangle \!\!\!&=&\!\!\!  \prod_{k=1}^{N} | + \rangle_{3k-2}
\prod_{k=1}^{N} \left[ \frac{1}{\sqrt{5}} \left( |2, -1 \rangle - |-1, 2 \rangle \right) 
- \sqrt{\frac{3}{10}} \left( |1, 0 \rangle - |0, 1 \rangle \right) \right]_{3k-1, \, 3k}. 
\nonumber
\end{eqnarray}
    
\end{enumerate}

\end{document}